\documentclass[fleqn,twoside]{article}
\usepackage{espcrc2}

\usepackage{graphicx}
\usepackage[figuresright]{rotating}
\title{ Light-front Hamiltonians for heavy quarks and gluons }

\author{ Stanis{\l}aw D. G{\l}azek 
         \address{ Institute of Theoretical Physics, Warsaw University, 
              ul.  Ho{\.z}a 69, 00-681 Warsaw, Poland.}
         \thanks{Work supported in part by MENiS BST-975/BW-1640}}
\begin{document}
%\preprint{IFT/15/05}
\begin{abstract}
A boost-invariant light-front Hamiltonian formulation of 
canonical quantum chromodynamics provides a heuristic 
picture of the binding mechanism for effective heavy
quarks and gluons.
\vspace{1pc}
\end{abstract}

% typeset front matter (including abstract)
\maketitle

\section{INTRODUCTION}

Light front (LF) Hamiltonian operators for effective particles in 
quantum field theory (QFT) provide a new path toward understanding 
of hadronic structure and interactions. The origin of new thrust 
is a far reaching simplification of the approach in comparison to 
the standard form of dynamics: boosts are kinematical, instantaneous
potentials can exist on the front hyperplane without contradicting
the assumption that no physical effect can propagate faster than 
light, and one can avoid the problem of the vacuum structure. 
Rotational symmetry is dynamical and poses problems, but we have 
a lot of intuition about rotational symmetry. It seems simpler to 
obtain rotational symmetry in the LF Hamiltonian approach than to 
solve the problem of dynamical boost symmetry or the vacuum 
problem in the standard approach. The price one has to pay for 
these gains in the LF Hamiltonian approach is a heavy-duty 
renormalization group procedure and a scheme to finesse a leading 
approximation around which one can develop a calculation of 
corrections. I describe a heuristic picture of binding for heavy 
quarks and gluons that begins to emerge from application of these 
tools in QCD. I will focus on results for gluonium and heavy quarkonia.

The binding of quarks and gluons occurs typically above 
threshold, which means that the sum of masses of the constituents 
is smaller than the mass of the bound state. How can such effect 
occur in a relativistic quantum theory? It does not happen in 
QED. So, how can it happen? A boost-invariant light-front Hamiltonian 
formulation of QCD provides a heuristic picture in which a binding 
between heavy effective quarks and gluons may arise above threshold 
because the eigenvalue condition for a renormalized Hamiltonian 
operator may include a large positive contribution from mass 
counterterms that can be compensated by exchange of gluons only 
in color-singlet states of limited size. One can conceive a 
subtle calculation without ever worrying about any non-trivial 
structure in the vacuum. But there is a trick needed to finesse 
effective interactions: one introduces an ansatz for a gluon mass 
gap in the Fock states that contain gluons in addition to the 
dominant sectors. Fortunately, the effective theory is not very 
sensitive to the ansatz and one can propose a scheme of successive 
approximations that may in principle replace the gap ansatz 
order by order with true interactions, ``true'' meaning implied by
the relevant theory. It is encouraging that even the first 
approximation exhibits a considerable degree of rotational 
symmetry. The key reference for this talk is \cite{ho}.

\section{KEY POINTS}

The framework for building a constituent picture for hadrons 
using LF Hamiltonians has several ingredients. The whole 
procedure begins with a canonical Lagrangian of QCD, but 
the actual Hamiltonian operators one works with are very 
different from the bare canonical ones.

1. In the momentum range where the binding mechanism works, 
the coupling constant $\alpha = g^2/(4 \pi)$ is comparable 
to 1. By this I mean that the relevant Hamiltonian contains
a coupling constant $g$ in its vertices, but the constant is 
a function of a renormalization group parameter $\lambda$.
When $\lambda \rightarrow \infty$, the constant $g$ tends
to 0 (it is never 0 for finite $\lambda$). In fact, the 
entire Hamiltonian as an operator is a function of $\lambda$,
\begin{eqnarray}
H  & = & H_\lambda \, .
\end{eqnarray}
The point is that one needs a method to find $H_\lambda$.
In QED, one can try to deduce a Hamiltonian from $S$-matrix 
theory, guessing potentials from scattering amplitudes within 
a perturbative expansion for very small coupling constant. 
In QCD, one expects confinement and if this expectation is 
taken seriously into account, including the large value of 
$\alpha$ in the binding mechanism, there is no precise link 
a la QED between the $S$-matrix for hadrons and quark and 
gluon forces responsible for the binding phenomenon. In 
order to develop LF Hamiltonians for quarks and gluons, I 
will use a renormalization group procedure for effective 
particles (RGPEP). The RGPEP provides an expression for
$H_\lambda$ order by order in an expansion in a small coupling
without reference to the $S$-matrix for quarks and gluons.
It also offers a possibility to extrapolate the small-coupling
results to large values of the coupling constant because the
extrapolation is for the Hamiltonian operator, not for the
observables (see points 5 and 6 below). But the Hamiltonian 
contains finite parts of the ultraviolet (UV) counterterms 
that are unknown, and the only possibility to find those 
unknown finite parts is to consider observables for hadrons, 
including symmetries such as rotational symmetry in decay 
amplitudes.

2. The Hamiltonians $H_\lambda$ must be clearly related
to the canonical theory. The RGPEP equations for evaluating 
$H_\lambda$ in perturbation theory (see e.g. \cite{ho}) are 
based on a unitary change of basis in the space of operators,
\begin{eqnarray}
q_\lambda & = & U_\lambda \, q_{can} \, U_\lambda^\dagger \, , \\
H_\lambda (q_\lambda) & = & \left[ H_{can} + H_{CT} \right]_{reg}(q_{can}) \, .
\end{eqnarray}
This is a similar idea to Gell-Mann's current-constituent 
relationship \cite{GellMann1,GellMann2} studied by Melosh 
\cite{Melosh1}, except that now it is built in a dynamical
scheme of QCD using RGPEP \cite{RGPEP}. The symbol $q_\lambda$ 
stands for creation or annihilation operators for effective 
quarks and gluons, and $q_{can}$ stands for canonical operators. 
$H_{can}$ denotes the canonical LF QCD Hamiltonian with a 
regularization. An admissible regularization must be imposed 
on the transverse and longitudinal relative momenta of 
interacting particles, those that are invariant with respect 
to the Poincar\'e transformations that do not take four-vectors 
out of the LF hyperplane $x^+=0$. $H_{CT}$ denotes the 
required counterterms. The RGPEP determines how to find the 
structure of $H_{CT}$ order by order in perturbation theory. 
The regularization and renormalization are ab initio in the 
Hamiltonian, defined en block for the operators in the Fock 
space that can be built using either $q_{can}$ or $q_\lambda$. 

\begin{figure}[ht]
\includegraphics[scale=0.37]{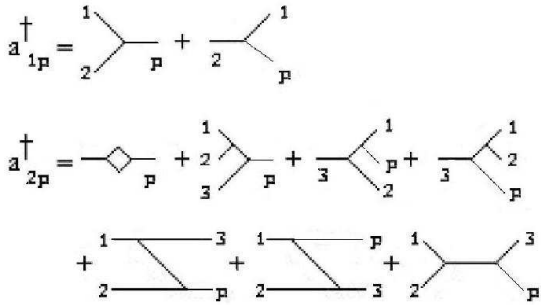}
\caption{ \label{adagger} \small
Structure of the first- and second-order terms 
in $a^\dagger_\lambda$ in a scalar theory of Ref. \cite{algebra}.
The thick lines on the left-hand sides of the vertices denote
canonical creation operators, and on the right-hand sides canonical
annihilation operators. The thin lines indicate how the momentum of 
an effective particle enters the structure.}
\end{figure}

3. In principle, a Fock-space decomposition of physical states 
contains wave functions that extend up to the cutoffs introduced
by regularization. Can a Hamiltonian formalism in a Fock space 
with not explicitly covariant regularization lead to covariant 
results? When Dirac introduced the front form of dynamics \cite{Dirac},
he reduced the problem to finding 10 generators that satisfy the 
Poincar\'e algebra:
\begin{eqnarray}
[P^\mu,P^\nu] &=& 0 \, , \\
{[}P^\mu,M^{\nu\rho}] &=& i(g^{\mu\nu}P^\rho - g^{\mu\rho}P^\nu) \, , \\
{[}M^{\mu\nu},M^{\rho\sigma}]
&=& i(g^{\mu\rho}M^{\sigma\nu}-g^{\mu\sigma}M^{\rho\nu} \nonumber \\
&+& g^{\nu\rho}M^{\mu\sigma} - g^{\nu\sigma}M^{\mu\rho}) \, . 
\end{eqnarray}
Can one seek such 10 generators in QFT using RGPEP? The seven
(one more than in the standard approach) kinematical generators: 
$P^+$, $P^\perp$, $M^{+-}$, $M^{+\perp}$, and $J^3$, do not 
require regularization and are the same as in a non-interacting 
theory. But the three dynamical generators: $H=P^-$ and $M^{-\perp}$, 
require regularization and renormalization. So far, it has only 
been demonstrated in a scalar theory up to second order in the 
coupling constant \cite{algebra} that RGPEP produces a required
solution. In that case one obtains commutation relations of the 
form ($g$ is the charge)
\begin{eqnarray}
[J^i_\lambda, J^j_\lambda] & = & i \epsilon_{ijk} J^k_\lambda + o(g^3) \, . 
\end{eqnarray}
There is no reason to expect that the method cannot work to higher 
orders. Dirac has observed that the challenge of an interacting 
theory is to unfold the commutators that contain products of
interactions. Such products appear in second-order expressions 
and a solution exists. Fig. \ref{adagger} illustrates the structure 
of the first- and second-order terms in the creation operators 
$a^\dagger_\lambda$ for effective particles. The zeroth-order term 
is $a^\dagger_{can}$ itself. 
The Dirac problem in QCD is much more complex than in a scalar
model theory. But it cannot be addressed in the RGPEP procedure 
without better understanding of the effective dynamics described 
by $H_{\lambda \, QCD}$ because of additional small-$x$ singularities 
in QCD that are not under control of the UV renormalization group 
procedure (see below).

4. A method for deriving Hamiltonians $H_\lambda$ must produce
operators that in the case of extremely small coupling constant
must deliver a covariant scattering matrix in perturbation theory. 
Wi\c eckowski \cite{dMarek} settled the issue in a 1-loop 
example of $\phi^3$ theory in 5+1 dimensions, which is asymptotically 
free (AF) in perturbative sense. His more general theorem states: 
The same $S$-matrix for scattering of physical particles can be obtained 
using (1) a bare Hamiltonian $\left[ H_{can} + H_{CT} \right]_{reg}$, 
and representing the in and out particles with creation and 
annihilation operators $q_{can}$, and (2) an effective Hamiltonian 
$H_\lambda$ and effective operators $q_\lambda$. In each order of 
perturbation theory, the result for the $S$-matrix is the same, 
provided that the relationship between $q_{can}$ and $q_\lambda$, 
and between initial $H$ and $H_\lambda$, is calculated up to this 
order. 

\begin{figure}[ht]
\includegraphics[scale=0.38]{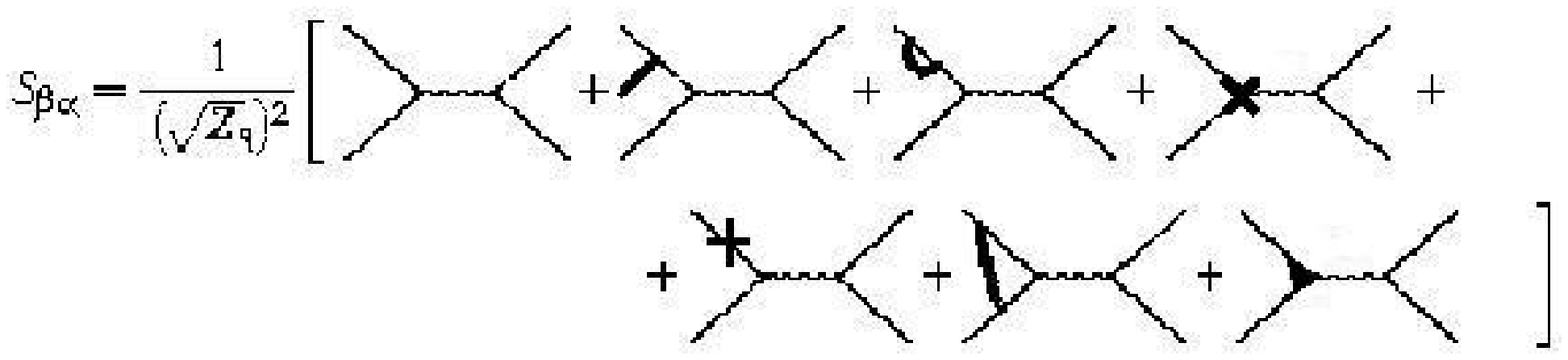}
\caption{\label{smatrix} \small Counterterms found using RGPEP 
produce a covariant amplitude of the type $ e^+e^- \rightarrow 
hadrons $, $S_{\beta \alpha}$ for transition from the incoming 
state $|\alpha \rangle$ to the outgoing state $|\beta \rangle$, 
in an old-fashioned one-loop calculation with two wave function 
renormalization factors $1/\sqrt{Z_q}$ in front \cite{dMarek}.}
\end{figure}

It is important that $H_\lambda$ for effective particles contains 
form factors in interaction terms. Thus, it resembles a non-local 
theory of the type known in relativistic nuclear physics, where 
nucleons interact through exchange of mesons. In distinction from 
the nuclear models, however, $H_\lambda$ originates from QFT and 
the form factors result from a renormalization group procedure. 
An example of old-fashioned diagrams that contribute to an amplitude 
of the type $ e^+e^- \rightarrow hadrons $ is shown in Fig. 
\ref{smatrix}.
Note that the integrals over momenta in the loops are limited in a 
non-covariant way because the momenta are already limited once and 
for all in a non-covariant way in the Hamiltonian. Nevertheless, 
the result for the scattering amplitude is the same as obtained 
using Feynman diagrams because the counterterms were found from
RGPEP and all they do is to remove the effects of regularization
\cite{dMarek}. This is interesting also from the point of view
of the world-sheet program developed by Thorn for planar diagrams
\cite{Thorn1,Thorn2}.

\begin{figure}[h]
\includegraphics[scale=.45]{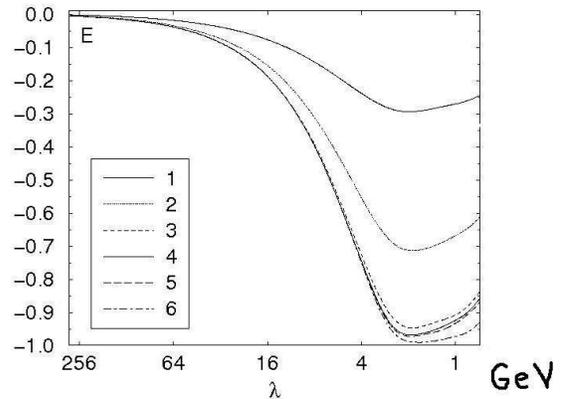}
\caption{\label{windowresult} \small A bound-state calculation
in an asymptotically free matrix model: a small window Hamiltonian 
matrix is evaluated in six successive orders of perturbation 
theory in RGPEP (using altered Wegner's flow equation) and then 
diagonalized non-perturbatively using a computer. The exact 
bound-state eigenvalue is -1 \cite{optimization2}.}
\end{figure}

5. How could one calculate bound state masses using $H_\lambda$?
This can be done like in atomic physics, where the Coulomb potential 
is only of formal order $e^2$ and still describes a giant variety of 
bound states. As soon as RGPEP produces $H_\lambda(g_\lambda)$, one 
can tackle the eigenvalue problem for $H_\lambda$. A special feature 
of RGPEP matters here: it works in perturbation theory without ever 
creating small energy denominators (this feature is built in the 
design of RGPEP following the principles formulated earlier in 
\cite{similarity1,similarity2}). Therefore, as long as $\lambda$ 
is above the size of momenta that matter in the binding mechanism,
the resulting Hamiltonians can provide relatively small matrices, 
called windows, whose diagonalization produces eigenvalues for the 
bound states. But the closer one approaches the scale of binding,
the higher order calculation is required and the larger is the 
required coupling constant $g_\lambda$. So, the window must be 
of the right size for the procedure to work. This aspect of the 
RGPEP scheme was carefully tested numerically in an exactly soluble  
model with AF \cite{optimization2}. A benchmark method of altered
Wegner's equation \cite{Wegner1,Wegner2,optimization1} produces 
Fig. \ref{windowresult} which shows how successive orders improve 
the accuracy of evaluation of the 
window in terms of its bound-state eigenvalue. Note that the 
parameter $\lambda$ must be near the scale of binding for a small 
window to work.
The model also indicates that a low order perturbative window,
obtained for arbitrarily small coupling (or arbitrarily small
$\Lambda_{QCD}$ in the RGPEP scheme) can be extrapolated to 
realistic values of $g_\lambda$ and then the window renders a
good approximation for the bound-state eigenvalue of the whole
theory. In future, after the boost-invariant RGPEP leads to 
identification of dominant terms via perturbation theory, 
the Wegner equation, or an altered version of it, can be applied 
to calculate window Hamiltonians for large coupling constants 
in specific cases without using perturbation theory. 

6. In the case of QCD, in the effective-particle basis in the 
Fock space, a gluonium state can be written as
\begin{eqnarray}
\Psi \rangle &=& 
|g_\lambda g_\lambda \rangle + |g_\lambda g_\lambda g_\lambda \rangle + ... \, ,
\end{eqnarray}
and a heavy quarkonium as
\begin{eqnarray}
\Psi \rangle &=& |Q_\lambda \bar Q_\lambda \rangle  
              + |Q_\lambda \bar Q_\lambda g_\lambda \rangle + ... \, .
\end{eqnarray}
But why should such expansion converge at all, especially when
the coupling constant is comparable with 1? The reason is that
the growth of $g_\lambda$ when $\lambda$ is lowered using RGPEP
can be compensated by the narrowing of form factors $f_\lambda$
in the interaction vertices. In other words, the interactions 
in a Hamiltonian with a small $\lambda$ die out so quickly for 
large energy changes that they are effectively not strong enough 
to spread probability to sectors with many effective particles, 
even when the coupling constant itself is large \cite{largep}.
The form factors of RGPEP solve the problem that the interactions
among effective constituents must be strong and at the same time
the number of the effective constituents must be small (as
indicated by the success of the constituent quark model). The 
situation is akin to nuclear physics with practically fixed 
number of nucleons.

7. The next key point is that in order to solve the eigenvalue
problem for $H_\lambda$ in QCD, one has to truncate it in the
number of effective constituents anyway. But in order to do 
it in a systematically improvable way, I assume that there 
exists a shift in the energy of gluons due to non-abelian 
potentials that come out from RGPEP in QCD (there are no such 
potentials coming out in QED, cf. \cite{positronium}) and that 
the shift can be approximated in the form of a mass gap function 
for gluons. A well-known operation $R$ \cite{R} produces then an 
effective Hamiltonian in the dominant sectors: two effective gluons 
in a gluonium, and a pair of effective quarks in a heavy quarkonium. 
Let me follow Mas{\l}owski's analysis of gluonium \cite{dTomek}, as 
an example of the application of RGPEP (the example is analogous
to the quarkonium case described in \cite{ho}, although gluons
demand a more advanced analysis because one cannot employ a 
non-relativistic approximation to begin with). Matrix elements 
of the effective Hamiltonian between states of two effective 
gluons $|g_\lambda g_\lambda \rangle$ labeled 1 and 2 with relative 
momentum $k$, are of the form 
\begin{eqnarray}
&&H_{kk'} 
 = (E_k  + E_{CT \, k})\delta_{kk'} + W_{kk'} \\
&& + {1 \over 2} \sum_q Y_{kq} 
\left[ {1 \over E_{k0} - E_{q\mu}} 
+      {1 \over E_{k'0} - E_{q\mu} } \right] Y_{qk'}, \nonumber 
\end{eqnarray}
where $E_k$ is an eigenvalue of kinetic energy operator 
$T_\lambda$, equal $ (k^{\perp \, 2} + m_\lambda^2)/(x_1x_2P^+)$, 
$m_\lambda$ is the effective mass of gluons in $T_\lambda$, 
$E_{CT}$ is the contribution of the mass counterterm for gluons, 
$W_{kk'}$ is the effective interaction generated by RGPEP,
the sum stands for summation over the effective three-gluon 
basis states that are coupled to the two-gluon basis states 
by the interaction term $Y$ in $H_\lambda$, and $E_{q\mu}$ is 
the LF energy of the three effective gluons, each of which has 
a mass gap ansatz $\mu^2$ in place of $m_\lambda^2$. The mass 
gap depends on the relative momenta of 3 gluons and can be 
described using a Hamiltonian term
\begin{eqnarray}
T_\mu
= {1 \over 3!}\sum_{123} 
\int [123] \sum_{i=1}^3 \frac{\mu_i^2(123)}{p^+_i} 
|123 \rangle \langle 123| \, ,
\end{eqnarray}
where $[123]$ means the LF measure in the integration over 
momenta of the three gluons and the sum extends over 
their polarizations (only two transverse ones).

8. The trick with the mass gap ansatz is to make it 
correctable order-by-order according to the following 
rule \cite{LFQCD}. One can write $H_\lambda = T + V$ 
and add and subtract the gap term $T_\mu$, changing nothing. 
But one is free to do it introducing a ratio of the coupling 
constant to its physically correct value for a $\lambda$ 
that one finds suitable to work with in this scheme:
\begin{eqnarray}
H_\lambda  & = & T + T_\mu 
+ \left[V - \left( { \alpha_\lambda \over \alpha_{physical} }\right)^2 
\, T_\mu \right] \, . 
\end{eqnarray}
For $\alpha_\lambda = \alpha_{physical}$, nothing is changed.
But in a perturbative expansion for small $\alpha_\lambda$,
the added term is large while the subtracted term is negligible.
This is how one can attempt to incorporate the non-perturbative 
effects in the three-body sector and couplings to sectors with 
more effective gluons in the first approximation. The effect
of $T_\mu$ in second-order calculation of $H_{kk'}$ will be 
replaced by the actual interactions in fourth-order RGPEP
calculation plus new small ansatz correction that will be 
correctable in higher order, and so on. The problem one solves
this way is that unless the sector with 3 gluons is separated 
by a gap from the sector with two gluons, it is not legitimate 
to use perturbation theory in order to account for the coupling
between the sectors. Most probably, in higher order analysis, 
the ansatz for a gap will be pushed away to sectors with more 
effective gluons. Nothing can be said for certain yet because
the 4th order calculation has not been completed. Note, however,
that if the mass gap ansatz approximates the true QCD interactions
well, there will be only a little change in the leading, approximate 
picture associated with transition from the case with the ansatz
term $T_\mu$ and without the interaction $V-T_\mu$ to the case 
with $V$ and without $T_\mu$. But in both cases one can consider
extrapolation of the window Hamiltonians in expansion in powers
of the coupling $\alpha_\lambda$ from arbitrarily small values 
to $\alpha_{physical}$ that is comparable with 1, since the 
RGPEP form factors prevent the interactions from blowing up 
at large momenta for large couplings.

9. How does the binding mechanism work? The mechanism
I am talking about can be considered an extension of the
seminal work of Lepage and Brodsky \cite{LB} to the domain
of small momentum transfers. They were working on exclusive
processes with a large 
\begin{figure}[htb]
\includegraphics[scale=.36]{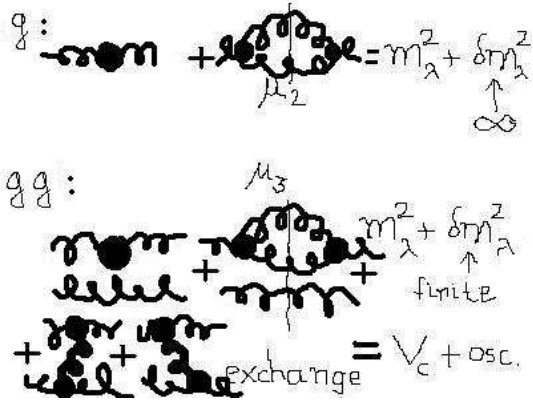}
\caption{\label{gluonbinding} \small The mechanism of binding
above threshold in gluonium. The upper line refers to how 
the mass counterterm for gluons is fixed using a single-gluon
eigenvalue equation for $H_\lambda$, and the lower line refers
to what happens in the eigenvalue equation for globally 
colorless states of two effective gluons, see the text.
The gluon lines are fat because they represent effective 
gluons.}
\end{figure} 
momentum transfer using diagrammatic 
rules for scattering amplitudes, while the RGPEP allows me 
to consider what happens in the Hamiltonian eigenvalue 
equation (with a gap ansatz) when the momentum transfer is 
in the range of the binding mechanism. Of course, Lepage and
Brodsky considered usual mesons and I discuss here gluonium,
but this is not essential and I will discuss a case of quarkonium
later. Let me explain how the binding above threshold emerges 
using Fig. \ref{gluonbinding}. This is the mechanism that
Mas{\l}owski used in his calculation \cite{dTomek}. 
The main difference in comparison with the earlier work
of Allen and Perry \cite{AllenPerry}, besides the explicitly
perturbative RGPEP which is different from coupling coherence
employed by Allen and Perry, is that RGPEP provides a 
boost-invariant eigenvalue equation for the mass of a 
gluonium in arbitrary motion, the basis states are built 
from effective gluons instead of the canonical ones, the 
form factors strongly limit changes of invariant masses
(instead of the changes of $P^-$ that are associated in 
\cite{AllenPerry} with interaction terms that depend on 
spectators), and one has to consider the coupling to 
three-gluon states in order to cancel small-$x$ divergences
(3-gluon states could be neglected entirely in \cite{AllenPerry}).  

First one considers an eigenvalue equation for a second-order 
$H_\lambda$ for states with quantum numbers of a single gluon,
see the first line in Fig. \ref{gluonbinding}. One assumes 
that the mass counterterm is such that if this eigenvalue
equation were solved in perturbation theory to second order in
$g_\lambda$ (first order in $\alpha_\lambda$) the physical gluon 
mass would come out equal 0. This condition produces an effective
gluon mass which is UV finite but contains a positive divergence
due to small-$x$ singularities (a blob on the single-gluon
line marked with letter g in Fig. \ref{gluonbinding}). The 
divergence is regulated in the initial LF QCD Hamiltonian. When 
one considers the same single-gluon eigenvalue problem beyond 
perturbation theory, and inserts a mass-gap in the two-gluon 
sector (now colored, different from the colorless states considered 
in the gluonium case below), the negative self-interaction 
(second term in the first line marked with g in Fig. 
\ref{gluonbinding}) is not able to work as strongly as in the 
case of perturbative, massless gluons, and the small-$x$
divergence in the effective gluon mass (coming from the counterterm)
is not compensated: the gap blocks the gluons from compensating it
because they cannot have the required small $x$ easily when they 
have a mass that vanishes too slowly for small $x$. As a result, 
a single gluon eigenstate may have an infinite mass in the limit 
of removing the small-$x$ regularization ($\delta m_\lambda^2$
in Fig. \ref{gluonbinding} diverges).

Next one considers the eigenvalue equation for colorless states
of two effective gluons and observes that the gluon mass gap 
in a colorless state may vanish much faster for small $x$ than
in states with color. Assuming that it is so, one can obtain 
a cancellation of the small-$x$ divergence and only a large 
self-interaction effect is left (finite $\delta m_\lambda^2$
in the lower line marked gg in Fig. \ref{gluonbinding}), 
sensitive to the behavior of the mass gap ansatz at small $x$. 
But there is also an exchange term which contributes also a finite
but large negative interaction when $x$ of the exchanged gluon is small.
This exchange term can compensate the large self-interaction effect,
but not entirely. There is a kind of quadratic potential well 
developing around a large minimum whose scale is related to the 
values of $\lambda$ and $\alpha_\lambda$ (marked as osc. in Fig.
\ref{gluonbinding}), in addition to a Coulomb force. Of course,
there are complex spin factors involved and it is a highly non-trivial
task to solve the equations numerically, see \cite{dTomek}. The bottom
line is that the renormalized self-interactions along the LF in 
$H_\lambda$ build up the gluonium mass high above threshold of 0.
The mechanism in heavy quarkonia is analogous but the analysis 
can be pushed analytically much farther because one can exploit the
non-relativistic limit when the quark masses are much larger than 
$\lambda$. This mechanism can be considered a generalization 
of the 1+1 dimensional model of 't Hooft \cite{tHooft} to
3+1 dimensional QCD.

\section{GLUONIUM}

\begin{figure}[h]
\includegraphics[scale=.35]{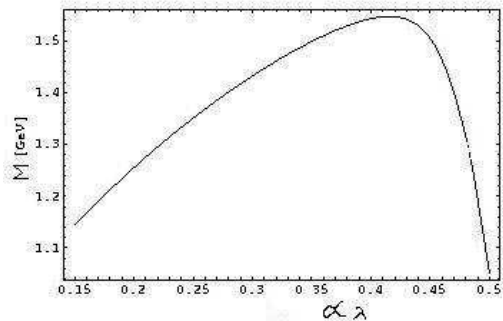}
\caption{ \label{run} Typical dependence of the smallest 
gluonium mass (in GeV) on the coupling constant $\alpha$
between 0.15 and 0.5.}
\end{figure}

In order to illustrate what comes out from LF Hamiltonians 
for effective gluons in pure gluodynamics, let me draw on
Ref. \cite{dTomek}. Mas{\l}owski \cite{dTomek} considered 
a class of the ansatz masses for every gluon in the 
three-gluon sector, $i = 1, 2, 3$, given by the same 
formula
\begin{eqnarray}
\label{ansatz}
\mu_i^2 & = & b^2 \, x_i \, (x_1 x_2 x_3)^\delta \, ,
\end{eqnarray}
where $b \sim \lambda \sim 2$ GeV,  and $\delta \sim 0.2$ 
are constants. A typical dependence of the smallest gluonium 
mass on $\alpha$ is shown in Fig. \ref{run}.
When the coupling constant increases, the state collapses.
But it is known from the beginning that the eigenvalue equation
with only two or three effective gluons cannot be valid for states
with masses much larger than two times an effective gluon mass.
The average value of the gap ansatz turns out to be about 
1.5 GeV in all interesting cases. Therefore, it is encouraging 
that there exists a range of couplings for which the gluonium 
mass is large and increases with the coupling. 

\begin{figure}[ht]
\includegraphics[scale=.35]{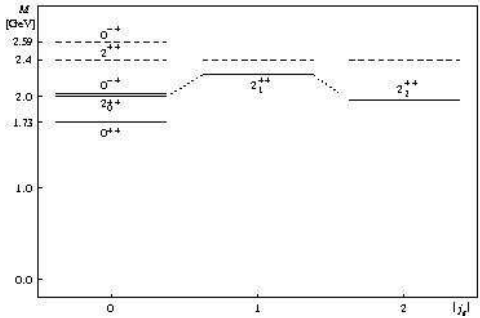}
\caption{ \label{gluonium1} \small 
Gluonium masses: dashed lines are taken from lattice
calculations \cite{lattice}. Bold lines are obtained
from $H_\lambda$. The mass of the state labeled $0^{++}$ 
was fitted to the lattice result of 1.73 GeV. The parameters 
in $H_\lambda$ are $\lambda = 1.92$ GeV, $b=2\lambda$, 
$\delta=0.2$, $\alpha_s=0.44$.}  
\end{figure}

An example of a spectrum of smallest masses from \cite{dTomek}
is shown in Fig. \ref{gluonium1} and in a table:
\vskip.1in
\begin{tabular}{|c|c|c|}
\hline
$M(j_z=0)$ & $M(|j_z|=1)$ & $M(|j_z|=2)$ \\ \hline
1.73 & 2.25 & 1.97 \\ \hline
2.01 & 2.63 & 2.54 \\ \hline
2.06 & 2.64 & 2.6  \\ \hline
2.41 & 2.66 & 2.67 \\ \hline
2.41 & 2.85 & 2.83 \\ \hline
\end{tabular}
\vskip.1in
These results represent a typical output from $H_\lambda$.
The example was obtained using 9 radial wave functions and
15 spherical harmonics. The results were stable with respect to
changes of these two numbers. The resulting spectrum is 
slightly
denser than on the lattice but otherwise it appears to be 
qualitatively similar. However, there is so far no clear way
for associating the eigenstates with spin quantum numbers. There
is no clear degeneracy into multiplets corresponding to
rotational symmetry. Although the lowest masses are not widely
different, and one may think that the violation of the multiplet
structure is not significant given the crude nature of the 
calculation (cf. \cite{AllenPerry}), the question of how to 
improve rotational symmetry is not answered. But it is certain
that a simple mass gap ansatz of Eq. (\ref{ansatz}) has no
a priori reason to produce true degeneracy of the masses.

Regarding the stability of the results versus changes of 
$\lambda$ one can say that when $\alpha_\lambda$ is forced 
to vary with $\lambda$ according to the perturbative formula 
with no quarks, the results are not varying significantly
with $\lambda$. But one can change the relative order of masses 
of $|j_z|=0, 1, 2$ eigenstates by making about 10\% changes 
of the values of $\lambda$, $\alpha$, $b$ and $\delta$. The 
overall conclusion is that the gluonium picture in the 
boost-invariant LF Hamiltonian approach appears surprisingly 
reasonable even in the crude first approximation. Readers
interested in more details should consult Ref. \cite{dTomek}.

\section{QUARKONIUM}

Theoretical aspects of the heavy quarkonium picture 
emerging from LF Hamiltonians for QCD with one heavy 
flavor and gluon mass gap ansatz are described in Ref. 
\cite{ho} and do not need to be repeated here. The 
main feature is that all details of the mass ansatz 
for effective gluons in the sector $|Q_\lambda \bar 
Q_\lambda g_\lambda \rangle$ disappear from the effective 
Hamiltonian in the $|Q_\lambda \bar Q_\lambda \rangle$ 
sector and the resulting second-order eigenvalue problem 
turns out to be exactly rotationally invariant. The 
first correction to the Coulomb potential is a force 
resembling a harmonic oscillator.

The fact that a mass gap ansatz leads to an oscillator-like 
interaction term which respects rotational symmetry already 
in the 2nd order analysis using $\mu^2 \sim 1$ does
not seem accidental. This result appears almost 
independently of all details of the ansatz because momentum
transfer squared, $q^2$, in the terms that are sensitive to 
small-$x$ of the gluons, is limited by the RGPEP form factors 
$f_\lambda$ to so small values that the ratio $\mu^2/(q^2 + 
\mu^2)$ is practically 1 for any reasonable ansatz. In addition, 
it seems likely that the same result comes out also as a 
part of the genuine 4th order calculation (not completed yet). 
In the 4th order, the ansatz term $\mu^2 \sim 1$ cancels out 
for large physical $\alpha$ \cite{ho}. The next term comes 
from QCD interactions order $\alpha$ in the 3-body sector and 
this is how the actual gap may show up. But $q^2$ in a system
dominated by the Coulomb force (the case of quark masses very
much larger than $\Lambda_{QCD}$) continues to be formally 
on the order of the strong Bohr momentum squared, or order
$\alpha^2$ which is much smaller than $\mu^2$ when $\mu^2 
\sim \alpha$. If this observation is confirmed in 4th order 
calculations, the spherical symmetry of the leading oscillator
term identified already in the 2nd order using the ansatz for 
$\mu^2$, may be a necessary consequence of the rotational
symmetry of QCD. 

One should observe two general arguments that support the 
harmonic result in the first approximation. One argument is 
that the combined effect of the self-interactions and the exchange
leads to binding above threshold which emerges around a minimum 
of the mass squared operator derived from the LF Hamiltonian 
$P^-$ and every function around a minimum is in the first 
approximation a quadratic one. The question is not so much 
why the potential comes out quadratic but rather what the 
spring (not string) constant is. Quite general Coulomb scaling 
argument implies that the spring constant is on the order of
$ k \sim \alpha \lambda^6 /m^3$. But if one assumes that
$\lambda \sim \sqrt{\alpha} m$, the resulting harmonic potential
scales with $\alpha$ exactly as the Coulomb eigenvalue problem
does. This means that such harmonic force may always be of a
fixed relative magnitude to the Coulomb part of the interaction
if QCD develops a mass gap for the effective gluons. The other 
argument is that when the distance between the heavy quarks 
increases, a linear potential may develop between the quarks
at the expense of creating additional gluons with a mass gap 
if the potential grows faster than linearly 
\cite{Wilsonparton,KogutSusskind} - it becomes energetically 
more favorable to create a definite number of gluons per unit 
of length than to allow the potential energy to grow without 
creating additional gluons. And the first possible integer 
power of the distance is 2. Thus, the harmonic force is 
probably the simplest one that can lead to a string picture
in LF Hamiltonians for QCD.

The simplicity of the LF Hamiltonian approach can now be 
exhibited using a calculation performed by a freshman 
on a personal computer \cite{Stawikowski}. When one 
neglects all spin effects and then disregards parameters 
that play no role in the spinless case, the eigenvalue 
equation for heavy quarkonia that one derives from the LF 
Hamiltonian reads
\begin{eqnarray}
\left[ -\frac{\Delta_r}{2m}\,-\,\frac{4 \alpha}{3 r} + \frac{k}{2}\, r^2 \right] \,
\Psi (\vec r) \, = \, E \, \Psi (\vec r) \, ,
\end{eqnarray}
where  $m$ is the reduced quark mass, equal half of the quark 
mass $m_q$. The eigenvalue $E$ gives the quarkonium mass through
the formula
\begin{eqnarray}
M = 2m_q \sqrt{1 + E/m_q } \, ,
\end{eqnarray}
which is relativistic.

Stawikowski found that for $k$ = 0.0431 GeV$^3$, $2 m = m_b$ = 5.0006 GeV, 
$\alpha$ = 0.6029, all compatible with the range of parameters a priori
possible in LF QCD, the masses of $\Upsilon$ states $1S$, $2S$, and $3S$, 
could be reproduced with precision of $10^{-5}$, or $\pm 0.05$ MeV. Such
accuracy is understandable because one has three parameters to fit three 
masses (Stawikowski considered a Coulomb potential multiplied by the factor 
$e^{-\mu r}(1-e^{\lambda r})$ but for the values $\mu=10^{-6}$ GeV and 
$\lambda=4$ GeV that he used, these additional parameters were not important).
The same parameters give $M_{1P}$ = 9.9144 GeV (the experimental average
of masses of $1P$ bottomonium states is 9.8884 GeV), and $M_{2P}$ = 10.2507 GeV 
(the experimental average of masses of $2P$ bottomonium states is 10.2519 GeV).

Keeping the same parameters $k$ and $\alpha$ and changing the quark mass to 
$m_c=1.6068$ GeV, one obtains the experimental value of the mass of $J/\Psi$ 
state, $M_{1S}$ = 3.0969 GeV. The $2S$ state obtains the mass $M_{2S}$ = 
3.6628 GeV (the experimental value is 3.6861 GeV), and the state $1P$ 
obtains $M_{1P}$ = 3.4991 GeV (the experimental average of masses of $1P$ 
charmonium states is 3.4940 GeV). It is clear that the harmonic oscillator 
potential as an admixture to the Coulomb potential is a qualitatively 
acceptable one from the point of view of such fit. An extensive numerical 
study including spin effects is under way \cite{heavy}.

\section{CONCLUSION}

The second-order RGPEP procedure for deriving LF Hamiltonians 
for heavy quarks and gluons produces operators that seem to 
have a chance of providing a qualitatively reasonable first 
approximation of the dynamical picture for masses of gluonium 
and quarkonia. Rotational symmetry requires intensive studies,
especially in the case of gluonium (important gluon degrees 
of freedom appear also in hybrids and there one faces considerable
constraints due to rotational symmetry \cite{hybrid}). The picture 
of binding mechanism that one can hold on to in order to build further 
intuitions is based on the possibility to control behavior of 
self-interactions of quarks and gluons in the effective Hamiltonians 
with small RGPEP parameter $\lambda$. The self-interactions largely
reduce positive contributions from counterterms, but not entirely,
and the exchange of gluons can lead to bound-state masses above 
threshold. The RGPEP procedure can be applied to many more cases 
than only those discussed here and one can refine the picture 
using higher order perturbation theory and a whole arsenal of 
methods for solving quantum Hamiltonian problems. An extension of
the approach to include light quarks will be more complicated
than the heavy quarkonium case. But the example of gluonium
suggests that the required fourth order RGPEP studies may shed 
new light on the issue of how to approach eigenvalue problems
including light quarks. So far, no vacuum structure was involved
in the entire calculation.

The rotationally symmetric effective Hamiltonians described here
provide an interesting addition to earlier studies based on 
similarity renormalization group and coupling coherence 
\cite{gcoh1,gcoh2,gcoh3,QQgcoh,QQ}. But it is not clear yet 
if the RGPEP procedure for LF Hamiltonians will ever match 
the accuracy of methods applicable in QED \cite{CasLep,krp} or
in lattice-based calculations \cite{nrqcd1,nrqcd2}. One has to be 
aware that a constituent picture can be fitted to data using a
large range of potentials \cite{MotykaZalewski,Zalewski}. 
On the other hand, if the hypothesis that QCD has an infrared
limit cycle \cite{irlcQCD,Wilsonirlc} is correct, it may turn out 
that new Hamiltonian methods provide a path to understanding of the
limit cycle universality \cite{universality} required for systematic 
solution of the theory. 

It is my pleasure to acknowledge discussions with Tomasz Mas{\l}owski, 
Jaros{\l}aw M{\l}ynik, Jakub Nar\c ebski, Lech Stawikowski, and 
Marek Wi\c eckowski. I thank Kenneth G. Wilson for numerous discussions 
on the renormalization of Hamiltonians at the early stages of the
development of the theory. I would like to thank the Organizers 
for a very interesting Workshop.

\end{document}